\documentstyle[aps,prl,twocolumn,floats,epsfig]{revtex} 
\begin{document}
 \wideabs{     
\title{Fourth-order magnetic anisotropy and tunnel splittings in Mn$_{12}$
from spin-orbit-vibron interactions} \author{Mark R. Pederson,$^a$  Noam
Bernstein$^a$ and Jens Kortus$^b$}
\address{$^a$Center for Computational Materials Science,
Code 6390, Naval Research Laboratory, Washington, DC 20375}
\address{$^b$MPI f\"ur Festk\"operforschung, Heisenbergstr. 1, D-70569 Stuttgart, Germany}
\date{\today}
\maketitle
\begin{abstract}
From density-functional-theory (DFT) based methods we calculate the vibrational
spectrum of the Mn$_{12}$O$_{12}$(COOH)$_{16}$(H$_2$O)$_4$ molecular
magnet. Calculated infrared intensities are in accord with experimental
studies. There have been no {\it ab initio} attempts at determining which
interactions account for the fourth-order anisotropy. We show that 
vibrationally induced distortions of the molecule 
contribute to the fourth-order anisotropy Hamiltonian and that the 
magnitude and sign of the effect (-6.2~K) is in good agreement with
all experiments. Vibrationally induced tunnel splittings in isotopically 
pure and natural samples are predicted.
\end{abstract}   

}

The possibility that
vibrationally induced modifications of the spin-orbit interaction
affects magnetic reorientation barriers has not been previously 
investigated. A
simple model captures the physics but quantitative determination
of the coupling constants requires computationally demanding
DFT methods.  A model Hamiltonian for a single uniaxial
anisotropic spin coupled to a one-dimensional harmonic oscillator is
given by $H = \gamma_{zz} S_z^2 + \frac{1}{2}(P^2+\omega^2 Q^2) + Q
\sum_{ab} \gamma_{ab}^{'} S_aS_b$ where $a$ and $b$ are summed over
$x$, $y$, and $z$, and $\gamma_{ab}^{'}=\frac{d\gamma_{ab}}{dQ}$).
Completing the square shows the diagonal
energy of a harmonic oscillator
($\left|\phi\right\rangle$) and  spin ($\left|SM\right\rangle$)
product state
as a function of the magnetic quantum number  ($M$) is given by:
\begin{equation} E= \omega/2 + \gamma_{zz} M^2 - (A+BM^2)^2/2\omega^2
\end{equation} with $A=S(S+1)(\gamma_{xx}^{'}+\gamma_{yy}^{'})/2$
and $B=\gamma_{zz}^{'}-(\gamma_{xx}^{'}+\gamma_{yy}^{'})/2$.
For this simple case, the interaction between vibrational and spin
degrees of freedom always acts to further stabilize each M state
but the energy splitting between the $|M|=S$ and $M=0$ states may
be either enhanced or reduced depending on the derivatives of the
$\gamma$ matrix. A more detailed analysis of this Hamiltonian, given
below, shows that it can also connect $M$ and $M \pm 4$ levels.
Further, in a real system,
the frequency, $\omega$, depends on nuclear masses so isotope effects
can lead to small changes in the barrier or, as discussed below,
larger tunnel splittings in the anisotropy Hamiltonian.  
Both the intrinsic 4th-order and isotope effects can mediate resonant 
tunneling of magnetization which is
of great interest from the standpoint of quantum mechanics at the
mesoscale.\cite{sess-1,Mn12,Fried,Sangre,barra,mettes}  
In addition,
the magnitudes of terms in the anisotropy Hamiltonian determine the
suitability of nanoscale particles for use in magnetic memory and
quantum-computing applications.~\cite{sun,leuenberger}
In general, purely electronic contributions to higher-order
magnetic anisotropy scale as $[1/(2c^2)]^4$ ($c =$ speed of light)
but the spin-orbit-vibron contribution scales as $[1/\omega^2]
\times [1/(2c^2)]^4$.  
The large $1/\omega^2$ prefactor suggests
this term could be dominant in high-symmetry bulk systems where
second-order effects vanish by symmetry.  

Recent experiments
on Mn$_{12}$O$_{12}$(COOR)$_{16}$(H$_2$O)$_2$
molecular nanomagnets, commonly referred to 
as Mn$_{12}$-Ac,~\cite{sess-1,Mn12,Fried,Sangre,barra,mettes,musfeldt} 
provide
an ideal system for understanding how vibrational degrees of freedom may
enhance the magnetic barriers. Many researchers have noted that
the tunneling dynamics may be mediated by conventional spin-phonon 
interactions.~\cite{garg} Further, recent experiments by Sushkov
{\em et al.}~\cite{musfeldt} show some strong variation in 
infrared (IR) spectra as 
a function of applied field suggesting a coupling between
the spin and vibrational degrees of freedom.
Accurate experiments have determined that
this easy-axis uniaxial molecular magnet has a second-order anisotropy 
parameter of $D=-0.56$~K and that the fourth-order 
contributions to the anisotropy Hamiltonian increase the barrier from 56~K
to approximately 65~K.~\cite{Fried,barra,mettes} 
Pederson and Khanna have combined DFT, spin-orbit coupling and second-order perturbation theory to 
calculate the second-order parameter (D) and find D=-0.557~K in
excellent agreement with experimental measurements.~\cite{ped1} 
An estimate of the entirely electronic spin-orbit-induced fourth-order 
anisotropy may be obtained by comparing the barrier computed with 
spin-orbit treated within second-order perturbation theory to that 
computed using exact diagonalization.
Within this approach we find a fourth-order contribution to the barrier 
on the order of 1~K and that it acts to reduce rather than to enhance 
the barrier.   

To determine if vibrational coupling is an important 
contribution
to magnetic anisotropies in Mn$_{12}$-Ac we have performed accurate
DFT based calculations on a 
single molecular unit. The
Perdew-Burke-Ernzerhof generalized-gradient approximation has been used
for all calculations.~\cite{pbe} Using
the methods discussed in Ref.~\cite{pore-vib}, 
the vibrational frequencies, vibrational eigenvectors, and IR and
Raman spectra have been calculated. Because this requires a large number (163)
of unsymmetrized molecular calculations we have used a 
Troullier-Martins type~\cite{trl-mar,por-pp}
pseudopotential for the O and C atoms and have treated the H and Mn atoms
within an all-electron method. Gaussian basis sets have been fully optimized
for each atom using the methods of Ref.~\cite{por-bas}.  
To determine the
anisotropy Hamiltonian and the  derivatives with respect to each normal     
mode, we calculate the anisotropy Hamiltonian $H=\sum_{ab}\gamma_{ab}S_aS_b$
for each of the 163 inequivalent vibrational displacements
using the method of Ref.~\cite{ped1}         
The coefficients $\gamma_{xy}$ for an arbitrary configuration of atoms
are determined from matrix elements of the spin-orbit-coupling operator 
sandwiched between all pairs of occupied and unoccupied Kohn-Sham 
wavefunctions (squared) and appropriate energy difference denominators.
The wavefunctions and thus the anisotropy Hamiltonian
depend on the geometry of the molecule.                                 
Derivatives of the anisotropy matrix with respect to the j$^{th}$ 
normal mode  
(i.e. $\frac{d\gamma_{ab}}{dQ_j}$) may be determined using a
finite-difference approach. 
We have 
ascertained that the partial-pseudopotential-based anisotropy Hamiltonian 
for the equilibrium geometry reproduces the all-electron anisotropy 
Hamiltonian of Ref.~\cite{ped1}. To test the accuracy of the
vibrational frequencies we compare our calculated
IR spectra, plotted in Fig.~1, directly to experimental measurements of 
Sushkov {\it et al.}\cite{musfeldt}

The experiments measure the IR absorption of Mn$_{12}$ crystals
suspended in paraffin pellets at wavenumbers ranging from 30~cm$^{-1}$
to 70~cm$^{-1}$ and from 140~cm$^{-1}$ to 650~cm$^{-1}$.  The
experimental absorption peak at 38~cm$^{-1}$ has a clear corresponding
feature in our calculations at 63~cm$^{-1}$.  The structure in
the 140~cm$^{-1}$ to 300~cm$^{-1}$ range is well reproduced by
the calculations:  The small experimental peaks at 150~cm$^{-1}$,
170~cm$^{-1}$, and 200~cm$^{-1}$ correspond to simulated features at
144~cm$^{-1}$, 170~cm$^{-1}$, and 201~cm$^{-1}$, respectively, and the
relative intensities are moderately well reproduced.  The small peak
in the simulation at 180~cm$^{-1}$ can be tentatively identified in
the experiment as a small peak at 185~cm$^{-1}$.  The structure
between 215~cm$^{-1}$ and 235~cm$^{-1}$ in the experiment, consisting
of a large peak, a small peak, and a large peak with a shoulder at
high frequency, has a clear analogue in the simulation results between
230~cm$^{-1}$ and 260~cm$^{-1}$ with similar relative intensities.
The intense peak at 255~cm$^{-1}$ corresponds to the simulation peak
at 275~cm$^{-1}$, with the smaller experimental peak at 270~cm$^{-1}$
appearing in the simulation as a shoulder at 260~cm$^{-1}$.  The double
peaked structure observed in experiment at 284~cm$^{-1}$, which
has significant activity in a magnetic field, corresponds to the
simulated peaks at 302~cm$^{-1}$, 313~cm$^{-1}$ and 316~cm$^{-1}$.
The following experimental triplet, consisting of a large peak
at 300~cm$^{-1}$ followed by two small peaks at 320~cm$^{-1}$
and 340~cm$^{-1}$, is present in the simulation, although at the
resolution of the graph in Fig.~1 the two peaks at 326~cm$^{-1}$ and
329~cm$^{-1}$ overlap, and the third peak appears at 343~cm$^{-1}$.
The small peak at 360~cm$^{-1}$ and intense peak at 375~cm$^{-1}$
are reproduced in the simulation with opposite relative intensities
(or interchanged frequency order) in the double peaked structure at
360~cm$^{-1}$ to 368~cm$^{-1}$.  The intense peak at 410~cm$^{-1}$
bracketed by two smaller peaks at 395~cm$^{-1}$ and 415~cm$^{-1}$
appears in the simulation between 412~cm$^{-1}$ and 439~cm$^{-1}$.
The two small peaks in the calculated IR spectrum at 385~cm$^{-1}$
and 395~cm$^{-1}$ are not clearly visible in the experimental data,
although they could correspond to the small feature at 382~cm$^{-1}$ to
390~cm$^{-1}$ in the gap between the prominent double peak and triple
peak structures.  In the highest frequency range measured by experiment
agreement with the simulation is still good, although the relative
intensities of the peaks are less accurately reproduced.  The two
faint experimental peaks at 465~cm$^{-1}$ and 495~cm$^{-1}$ probably
correspond to the two relatively prominent peaks at 460~cm$^{-1}$ and
476~cm$^{-1}$.  The five peaks between 510~cm$^{-1}$ and 570~cm$^{-1}$
correspond to the five peaks between 500~cm$^{-1}$ and 545~cm$^{-1}$.
Although the intensities relative to the previous two peaks are lower
in the simulation, opposite to experiment, the relative intensities
within the five peak structure are good.  In both the experiment and
simulation the second and fifth peaks are most prominent.  The intense
double peak at 605~cm$^{-1}$ to 610~cm$^{-1}$ clearly corresponds to
the simulated double peak at 564~cm$^{-1}$ to 571~cm$^{-1}$.  There is
no comparably intense feature in the simulation that corresponds to
the highest frequency experimental peak at 640~cm$^{-1}$, although
several less intense peaks are present in the correct frequency range.

The decomposition into contributions from different structural
elements in the Mn$_{12}$ molecule shows some general trends as a
function of frequency, but no sharp variation from mode to mode.
Below 500~cm$^{-1}$, the main contributions are from the Mn and
formic acid.  The water molecules contribute mainly for modes below
about 350~cm$^{-1}$, while the anionic oxygen atoms contribute mainly
above this frequency.  Above 500~cm$^{-1}$ the Mn contribution grows
relative to the formic acid contribution, with particularly large
Mn and anionic oxygen weights in the 564~cm$^{-1}$ to 571~cm$^{-1}$
double peak.

We now discuss our calculations on the vibrational contributions to 
fourth-order anisotropy.  The coupled spin/vibron Hamiltonian is given
by:
\begin{eqnarray}
H & = &\sum_{ab}\gamma_{ab}S_aS_b \\ \nonumber
&&+ \sum_{j}\frac{1}{2}\left[P_j^2+\omega_j^2\left(Q_j^2 
+ \frac{2}{\omega_j^2}\sum_{ab}\frac{d\gamma_{ab}}{dQ_j}Q_jS_aS_b\right)\right]. 
\end{eqnarray}
The above expression may be analyzed either classically or 
quantum-mechanically to determine how vibrational coupling affects 
the magnetization barrier. 
As demonstrated below 
the magnetization barrier is enhanced by approximately 5.6~K classically
and 6.2~K quantum mechanically since the quantum mechanical 
expection value of 
S$^2$ is 10 \% larger than the classical value for S=10.
However, to address questions related to isotopic effects
and changes in tunnel barriers a quantum mechanical analysis is required.  

The simplest level of approximation
is to take a product wavefunction of the form 
$\left|\Psi\right\rangle = \Pi_j\left|\phi_j\right\rangle\left|SM\right\rangle$,
with $\left|\phi_j\right\rangle$ a harmonic oscillator wavefunction and $\left|SM\right\rangle$ a   
spin-wavefunction, and determine the diagonal energy of this state.
The resulting energy $E(M)=\left\langle SM \right|H\left| SM \right\rangle]$ is given by:
\begin{eqnarray}
E(M) & = & \sum_{ab} \gamma_{ab} U^{ab}(M) 
      + \sum_{j} \frac{1}{2}\omega_j \\ \nonumber
  & & - \sum_{abcd} \Gamma_{abcd} U^{ab}(M)U^{cd}(M)
\end{eqnarray}
with $\Gamma_{aba'b'}=\sum_j\frac{1}{2\omega_j^2}\frac{d \gamma_{ab}}{dQ_j} 
\frac{d \gamma_{a'b'}}{dQ_j}.$ and U$^{ab} =\left\langle SM \right|S_aS_b\left| SM \right\rangle]$ 
The 4th-order barrier enhancement
may then be estimated by evaluating the second line of Eq.~(3) for
the $M = \pm 10$ and $M=0$ states and taking the difference. This
energy difference is 6.2~K.

To compare directly to the experimental parametrization of Barra {\em et
al.} which uses 4$^{th}$-order Stevens operators, it is instructive              to expand both our calculated and their fitted  
4th-order expression in terms of orthogonal cubic polynomials
of degree 4. 
Because of the S$_4$ symmetry of the molecule, 
it is generally possible to write the fourth-order energy according to:
$E^4 = A_o^4 + (
  A_1(4)[S^2(S_z^2-S^2/3)]
+ A_2(4)[3S^4+35S_z^4-30S^2S_z^2]
+ B_1(4)[S_x^4+S_y^4-6S_x^2S_y^2]
+ B_2(4)[S_xS_y(S_x^2-S_y^2)])/10^4$.
In Table I,
we compare the values of the expansion coefficients for each 
representation. The O$_4^2$ Stevens operator can be expanded into 
Legendre polynomials of degree 2 and 4 respectively. 

An interesting feature, present in both the experimental
and theoretical expansions, is that for the most part, the
4$^{th}$ order terms exhibits nearly the same angular variation as
the second-order Hamiltonian. For example 
the entirely diagonal 4$^{th}$ order $S^2(S_z^2-S^2/3)$ term
accounts for almost
all of the barrier enhancement.
This
feature is very important from the standpoint of sharp tunneling 
transitions. It ensures that magnetic-field induced alignment 
condition aligns multiple pairs of states simultaneously which 
opens multiple
tunneling paths. The coefficient $A_2(4)$ changes the diagonal 
energies by $A_2(4)\times M^4/10^4$ and is partially responsible
for a broadening of the magnetic-field alignment condition but
does not cause any tunnel splittings at zero-field. The $B_1(4)$
coefficient corresponds to the $B^4_4$ Stephenson coefficient,
and leads to observable tunnel splittings in  
the $M=\pm2$ manifold and smaller tunnel splittings in the $M=\pm4$ manifold.
Our isotopically pure calculations show tunnel splittings of the 
$M=\pm2$, $M=\pm4$, and  $M\pm=6$  manifolds of approximately
10$^{-2}$, $7\times 10^{-6}$ and  $6 \times 10^{-10}$ K respectively
but do not split the odd M states since coupling is between M and
M$\pm 4$. By randomly changing a single atomic mass by 1 amu, we
find that the odd-M states are split with a tunnel splitting of the
$M=\pm 1$ states on the order of 10$^{-4}$ K. The isotopic  
effects  experimentally observed by Wernsdorfer et al.~\cite{werns} in
Fe$_8$ may be partially due to the mass effects identified here.

\begin{figure}
\begin{center}
\epsfig{file=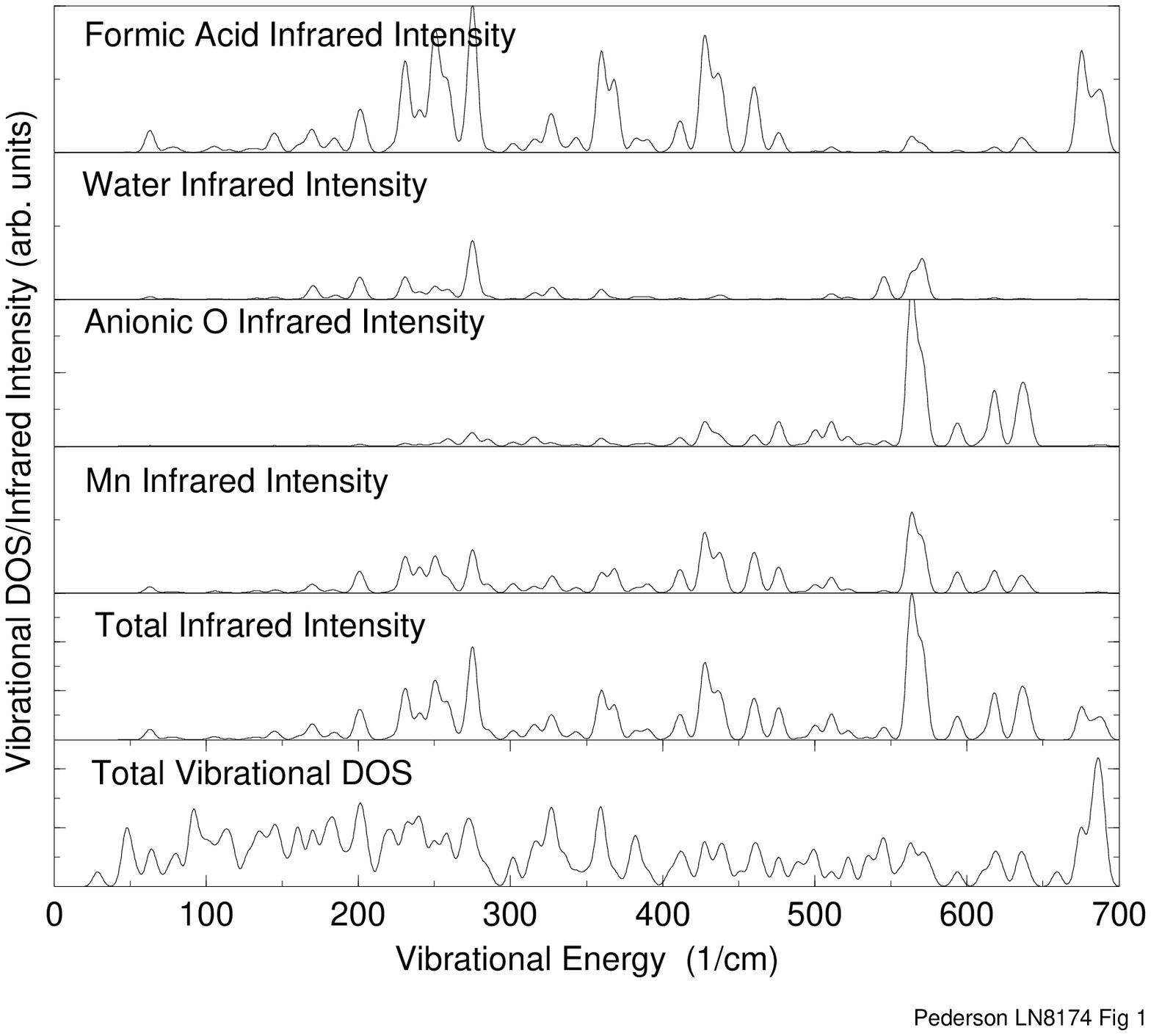,width=3.0in,clip=true}
\end{center}
\caption{Calculated total and IR vibrational density of 
states for the Mn$_{12}$ molecule. In addition we have projected
the IR active density of states onto Mn, O$^{2-}$, COOH, and H$_2$O to
show the origin of the IR spectrum.  }
\label{fig1}
\end{figure}          
\begin{figure}
\begin{center}
\epsfig{file=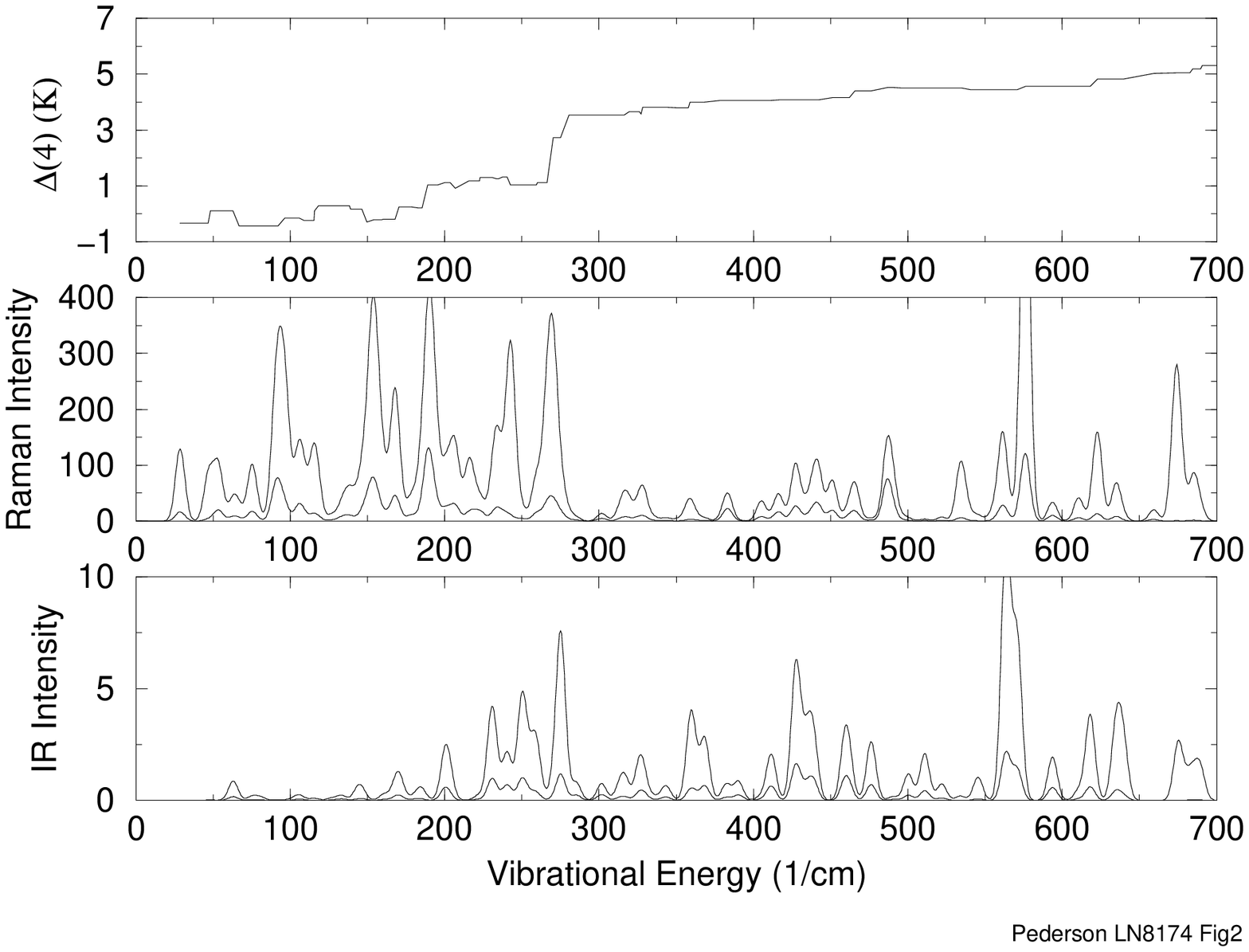,width=3.0in,clip=true}
\end{center}
\caption{Evolution of 4th order anisotropy barrier ($\Delta(4)$)
as a function of the number of modes included. Also shown
is the total and Mn-projected calculated IR and Raman spectra. 
Large jumps in the barrier are due to
strong Raman modes, as discussed in the text. }
\label{fig2}
\end{figure}          

In Fig.~2 we show the barrier as a function of the
number of vibrational modes that are included. When none of the
vibrational modes are included the barrier reproduces the
earlier calculation of Ref.~\cite{ped1}. As expected qualitatively,
most (85 \%) of the fourth-order barrier is associated with
the frequencies in the 100--500 cm$^{-1}$ range that correlates with
Mn vibrations. The upper panel of the figure  shows
that there are a 5--10 modes which account for about 80--90~\% 
of the fourth-order anisotropy barrier. For example, single
modes at 189, 270, 280, and 465 and 1496 cm$^{-1}$ contribute 
0.8~K, 1.6~K, 0.8~K, 0.25~K and 0.45~K respectively to the 4th-order
barrier. These modes and all other modes that contribute
visible spikes in Fig.~2 share one simple
trait. They  are optically silent to IR 
absorption and transmission but show strong Raman 
intensity. The calculated Raman intensities
for these modes are shown in Table II.

Recently, Sushkov {\em et al.}~\cite{musfeldt} have performed experiments
on Mn$_{12}$-Ac and have shown that the IR transmittances
at 284, 306 and 409 cm$^{-1}$
exhibit strong dependencies when magnetic fields are 
applied to the crystal and that the dependence is particularly
strong for the mode at 284 cm$^{-1}$.  This 
suggests that vibrations in this energy range are associated
with the spin carrying Mn ions as confirmed from the
projected DOS in Fig.~1. While the
Raman modes are responsible for the formation of the 4th-order
barrier, the strong contributions due to Mn motion at 270 and
280 cm$^{-1}$ is qualitatively consistent with the 
strong field-dependent IR dependencies observed experimentally in
this energy range.

\begin{table}[bh!]
\caption{Calculated Raman intensities for
modes that are primarily responsible for the formation
of the vibrational 4th-order magnetic anisotropy. Units
of Raman intensities are respectively $\AA^4$/amu.}
\begin{tabular}{lrr}
$\omega$ (cm$^{-1}$) &   Raman   & 4th-order shift(K) \\
189            & 297     &      0.8        \\
270            & 508     &      1.6        \\
281            &  24     &      0.8        \\
465            & 263     &      0.25       \\
1496           &5400     &      0.45       \\
\end{tabular}
\end{table}
\begin{table}[bh!]
\caption{Fourth-order anisotropy Hamiltonian as determined 
from experiment, DFT plus vibration-spin
coupling, and DFT without this coupling.
For simplicity we have reexpanded the 
representation of Barra {\em et al.} in terms of orthogonal
cubic polynomials rather than the Stephens polynomials used in
their work. All numbers are given in K$\times10^4$.}
\begin{tabular}{lrrrr}
           & A$_1$(4) & A$_2$(4) & B$_1$(4)&B$_2$(4) \\
Experiment    & -8.35  & -0.334 &-0.43  &  0.000\\
Vibrational   & -5.58  & -0.008 &-0.01  & -0.015\\ 
Electronic    &  0.68  &  0.0005&-0.002 &  0.004\\
\end{tabular}
\end{table}

While the calculated 4$^{th}$-order contribution to the
tunneling barrier of 6.2~K is in close agreement with 
the experimental values of  7--10~K, it is indeed a small
number so it is appropriate to consider
other vibrational effects that might be nonnegligible.

To determine possible effects of methyl termination of the carboxyl
groups, we have changed the mass of the H on the formate groups
from 1 to 15 (the mass of a CH$_3$).  This further increases the 4th-order 
anisotropy from 6.2~K to 7.3~K  
and leads to even better agreement
with experiment.
There could also be  terms due to 
d$^2 \gamma /dQ^2$.  However, such interactions
effectively change the vibrational spring constant
and scale as $\frac{\mu^2}{M \omega c^4}$ 
rather than $\frac{\mu^4}{\omega^2 c^8}$ where $\mu$     
is the total magnetic moment. Because of the large masses 
involved (even for hydrogen) an order of magnitude estimate
suggests that the total barrier would change by less than 0.01~K
from this type of effect. Further, in contrast to terms discussed
above, these contributions 
add as amplitudes and could partially
cancel one another.  The spring-constant 
terms directly modify the 2nd-order $\gamma$ matrix allowing
for isotopic induced symmetry breaking. While
potentially unimportant from the standpoint of barrier formation,
this effect leads to a type of
{\it locally varying second-order transverse anisotropy}
required to explain experiments (See Ref.~\cite{mertes}.)
We are in the process of studying this and other possibilities.

To summarize, we have performed accurate DFT-based
calculations on the Mn$_{12}$-Ac molecule to determine whether
vibron-spin coupling could be responsible for part of the
4th-order anisotropy Hamiltonian. Our results suggest that 
vibron-spin coupling accounts for most of the effect and
that it is significantly larger and a different sign than the O(1/c$^8$) terms
that arise from an exact diagonalization treatment of spin-orbit coupling.

This work was supported in part by ONR grant N0001400WX20111 and
N0001401WX31303.

\end{document}